# Reconstruction-Stabilized Epitaxy of $LaCoO_3/SrTiO_3$(111) Heterostructures by Pulsed Laser Deposition

Minhui Hu[1], Qinghua Zhang[1], Lin Gu[1,2,3], Qinlin Guo[1], Yanwei Cao[4], M. Kareev[4], J. Chakhalian[4], JiandongGuo[1,2,3]

1. Beijing National Laboratory for Condensed Matter Physics & Institute of Physics, Chinese Academy of Sciences, Beijing 100190, China.

2. School of Physical Sciences, University of Chinese Academy of Sciences, Beijing 100049, China.

3. Collaborative Innovation Center of Quantum Matter, Beijing 100871, China.

4. Department of Physics and Astronomy, Rutgers University, Piscataway, NJ 08854, USA.

**Abstract**

Unlike widely explored complex oxide heterostructures grown along [001], the study of [111]-oriented heterointerfaces are very limited thus far. One of the main challenges is to overcome the polar discontinuity that hinders the epitaxy of atomically sharp interfaces. Here, by taking the $LaCoO_3/SrTiO_3$(111) as a prototype, we show that the reconstruction, which effectively compensates the surface polarity, can stabilize the epitaxy of the heterostructure with polar discontinuity. Reconstructed substrate surface is prepared, while the growth is controlled to form reconstruction on the film surface. To suppress the chemical diffusion across the interface, the growth is interrupted between each unit cell layer to allow the lattice relaxation at a lowered temperature. In this way, high quality two-dimensional growth is realized and the heterointerfaces exhibit sharpness at the atomic scale. Our work provides a path to precisely control the growth of complex oxide heterostructures along polar orientations that exhibit emergent quantum phenomena.

Recently, due to the emergence of variously intriguing quantum phenomena, artificial graphene-like structures of complex oxides have attracted a great number of studies [1-7]. It has been pointed out that the (111) bilayer heterostructure of perovskite transition metal oxides (TMOs) has a natively inverted band structure resulting from the geometry of the honeycomb lattice [1]. Therefore, thin films, particularly bilayer heterostructures grown along [111] provide an ideal platform for realizing topological phases [1-4]. For example, $LaAlO_3/LaAuO_3/LaAlO_3$(111) heterostructure was proposed as a candidate to realize the quantum spin Hall effect [1], whereas the [111]-oriented $(LaNiO_3)_2/(LaAlO_3)_N$ heterostructure was predicted as a system with quantum anomalous Hall effect (QAHE) [3-4]. Compared to the previously found topological compounds [5-7], (111) bilayer heterostructures of TMOs with strongly correlated electrons have many advantages, including the chemical inertness, which paves an exciting path to realize topological phases in condensed matter systems by materials design.

Experimentally the key to realize topological phases in [111]-oriented TMO heterostructures is the control of sharp heterointerfaces at the atomic scale. To obtain the atomically flat substrate surface and to keep the two-dimensional (2D) growth mode of the epitaxy (i.e., layer-by-layer or step-flow), the net charges in each perovskite (111) atomic plane have to be compensated. The surface/interface reconstruction is one of the way for such compensation. The formation of complex surface reconstructions is normally effective in the [001]-oriented growth [8–11]. However, this requires high temperature growth to activate the cation relaxation, which also enhances the chemical interdiffusion at the heterointerfaces and degrades the sharpness. Especially for heterostructures with polar discontinuity, there would be an additional driving force for the intermixing of cations with different valences. Despite the growth of high quality [001]-oriented TMO heterointerfaces, the preparation of (111) heterostructures is still very challenging [12-14]. As the result, the experimental investigations have not provided conclusive evidence for the presence of predicted exotic quantum phases.

To address this challenge, we investigate polar heterojunction LaCoO$_3$/SrTiO$_3$ as a prototype to show the importance of surface reconstructions to compensate for the polar discontinuity and maintain the atomically precise growth. As shown in Fig. 1 (a), the [111]-oriented LaCoO$_3$/SrTiO$_3$ (LCO/STO) interface exhibits polar discontinuity. Due to its rich physical properties, LaCoO$_3$ (LCO) was subjected to numerous investigations both experimentally and theoretically. Bulk LCO, where the nominal oxidation state of Co is 3+ with a 3d$^6$ electronic configuration undergoes temperature-induced insulator-to-metal with a complex evolution of its anomalous spin-state [15-18]. Specifically, at low temperature, LCO is a nonmagnetic semiconductor with low-spin configuration ($t_{2g}^6$; S=0). With increasing the temperature to 100 K, it forms a paramagnetic intermediate spin state ($t_{2g}^5 e_g^1$; S = 1) and then becomes a high-spin state ($t_{2g}^4 e_g^2$; S = 2) above 500 K. Furthermore, very recently it was predicated that two LCO monolayers sandwiched between band insulator layers of STO along (111)-orientation can host topological phases and QAHE [19]. This is because the interplay between spin-orbit coupling and Coulomb interaction can stabilize a robust ferromagnetic insulating phase with a nonzero Chern number.

In this Letter, we explore the growth of perovskite LCO thin films and LCO/STO superlattices along the [111] direction. By preparing the reconstructed STO(111) substrate surface and keeping the film surface reconstructed, high-quality layer-by-layer growth is achieved by pulsed laser deposition (PLD). X-ray diffraction (XRD) confirms the [111] growth direction of the heterostructures without impurity phases. Detailed film thickness and interfacial roughness were investigated by X-ray reflectivity (XRR). The chemical stoichiometry and core level electronic structures of LCO thin films were characterized by X-ray photoemission spectroscopy (XPS) and synchrotron based resonant X-ray absorption spectroscopy (XAS). High-resolution scanning transmission electron microscopy (STEM) shows the formation of sharp interfaces between LCO and STO layers. All the measurements corroborate that the high quality quasi-2D (111) heterostructures can be achieved by precise control at the atomic scale.

The LCO thin films and LCO/STO superlattices were grown with a PLD system equipped with high energy electron diffraction (RHEED) using a KrF excimer laser operating at λ=248 nm. Both STO and Nb-doped (0.5 wt.%) STO(111) substrates of 2 × 10 mm$^2$, one-side polished and a 0.1° miscut were used; No significant doping effect of the substrate on the growth was detected. In the following we present the results on Nb-doped STO substrates except for the electric transport measurement. The thin films and superlattices were prepared using stoichiometric LCO and STO ceramic targets (99.9% purity). During the growth, the substrate was resistively heated to 700 °C in oxygen with the partial pressure of $5 \times 10^{-6}$ mbar. A KrF excimer laser with the energy density of 2.5 J/cm$^2$ and the pulse frequency of 10 Hz was used. All samples were post annealed in oxygen at 300 °C with the pressure of $5 \times 10^{-1}$ mbar for 1 hour, and then cooled down to room temperature at a rate of 15 °C/min. Both the prepared STO substrate and LCO film surfaces were characterized by *ex situ* atomic force microscope (AFM). The sheet resistance measurement of the LCO thin film was measured by PPMS (Quantum Design) in van-der-Pauw geometry at 195-310 K. The XRD and XRR measurements were performed with a Rigaku SmartLab (9 kW) X-ray diffractometer with Ge (220) × 2 crystal monochromator. The XPS measurements were carried out by the Thermo Scientific ESCALAB 250 spectrometer with a monochromated X-ray source (AlK$_\alpha$). The energy resolution was set at 0.5 eV as calibrated by the full width at half maximum (FWHM) of pure metal Ag 3d$_{5/2}$ peak. The photoelectrons were collected perpendicularly to the sample surface. The XAS measurement at Co L$_{2,3}$-edge were performed with a total electron yield (TEY) detection mode at Beamline 6.3.1 of the Advanced Light Source (ALS, Lawrence Berkeley National Laboratory). The grazing angle of incident X-ray beam was kept at 20°. The cross-sectional STEM characterizations were performed on an ARM-200CF (JEOL) operated at 200 keV and equipped with double spherical aberration (Cs) correctors for both probe-forming and imaging lenses. The attainable resolution of the probe was 78 pm as defined by the objective pre-field.

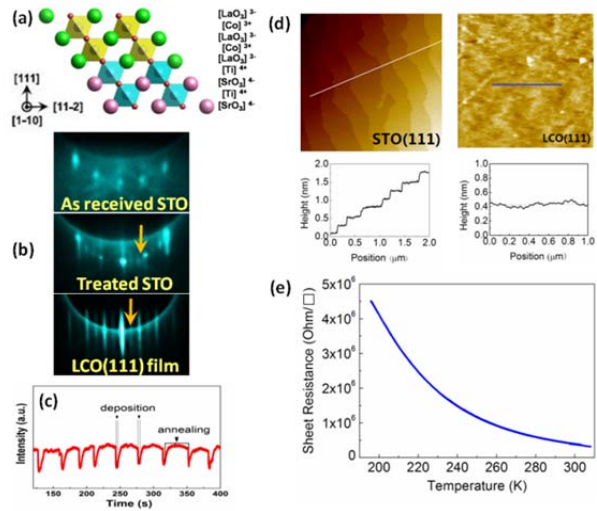

Figure 1 (a) Crystal structure and schematic of polar discontinuity of LCO/STO(111) heterointerface. (b) *In-situ* RHEED patterns of *as received* STO(111) substrate, treated STO(111) and LCO(111) film. All RHEED patterns have been recorded along the $[1\bar{2}1]$ direction of the samples. The arrows indicate the 5/6 and 1/2 fractional patterns of the reconstructions, respectively. (c) Evolution of the RHEED intensity during the interrupted deposition of LCO(111) film. Only a part of the growth is shown. (d) AFM images (2 μm × 2 μm) and according line profiles of both the treated STO(111) substrate and LCO film surfaces. (e) Sheet resistance of the LCO thin film as a function of temperature, exhibiting highly insulating characteristics, which is consistent with the previous reported [20].

We find two factors are critical to keep the stable, high quality 2D growth of the epitaxial films on reconstructions. One is to prepare the atomically flat substrate surface in reconstruction. There have been studies reported on the reconstructions of STO(111) surface after *in situ* treatments [21-22], which effectively compensate the surface polarity and allow the formation of atomically flat terraces separated by single-unit-cell steps. By adjusting the preparation parameters, the surface termination type as well as the reconstruction periodicity can be tuned [23-25]. In the current work, prior to the film deposition, we ultrasonically cleaned the substrates with acetone and ethyl alcohol, followed by *in-situ* annealing in oxygen with the pressure of $5 \times 10^{-6}$ mbar at 900 °C for 1 hour. The $Ti^{4+}$-terminated surface was obtained with the (6×6) reconstruction. As illustrated in Fig. 1(b), the streak patterns observed along the Laue circles together with the Kikuchi lines and the reconstructions confirm the flatness and good crystallinity of the substrate surface. The *ex situ* AFM image also demonstrates the atomically flat terraces with single-unit-cell-high steps [see Fig. 1 (d)].

The other critical growth factor is to maintain the surface reconstruction on the film. By such means, homoepitaxy of high quality STO polar films have been obtained in a 2D mode by molecular beam epitaxy [26-28]. However, a required high growth temperature (> 800 °C) generally limits the application of the method in heteroepitaxial growth due to the enhanced chemical interdiffusion. Here we introduce the interrupted deposition method [29-30] to synthesize high quality LCO films and LCO/STO superlattices at a lowered temperature. Relatively high laser pulse frequency of 10 Hz was used to deposit one unit cell (u.c.) in a short time interval ~ 2.8 seconds, followed by a much longer interval (~ 30 seconds) for the deposited film rearranged by annealing, as illustrated in Fig. 1 (c). Once ~ 1 u.c. LCO film was deposited onto the substrate, the initial "6×" fractional diffraction patterns from the STO substrate disappeared and the RHEED intensity decreased dramatically. Next the laser pulse was interrupted and the sample was annealed at the growth temperature in oxygen. Following this procedure, the RHEED intensity increased again until it recovered the previous value, and "2×" reconstruction patterns appeared. Note that the energy barrier for lateral lattice relaxation is normally lower than that of atomic diffusion perpendicular to the interface because the later involves the site exchange at a longer range (typically on the scale of a u.c.). Therefore, there is the temperature window in which the surface reconstruction can be formed by extended annealing during the interrupted interval, while the intermixing at the heterointerface can be suppressed to maintain its sharpness and high LCO film crystallinity. Based on this, we lower the growth temperature down to 700 °C while keeping the formation of growing layer reconstruction intact, *i.e.*, 700 °C is the optimized growth temperature. Such a growth cycle was repeated, and the LCO polar film was grown in high quality 2D mode. After deposition, the sample was cooled down to room temperature, and rotated by 60° relative to the original position to confirm the presence of the (2×2) reconstruction on the flat LCO surface. The *ex situ* AFM image shows a single terrace of the 10-nm-thick LCO film with the corrugation of ~ 0.1 nm [see Fig. 1 (d)], significantly lower than a u.c. height, and the broadened terrace evidences the high quality 2D growth. Note, on the *as-received* STO(111) substrate without reconstruction [Fig. 1 (a)], LCO immediately falls into the three-dimensional mode with high surface roughness.

The crystalline quality of the LCO(111) thin film on STO was investigated by XRD with 2θ/ω scans. As shown in Fig. 2 (a), the sharp XRD peaks indicate that the LCO film is a [111]-oriented single crystal without any detectable impurity phases. The c-axis lattice spacing of the LCO film is $d_{film}$=2.171 Å which agrees with the expected values for the fully strained lattice. The fringes corresponding to the LCO(111) reflection are clearly visible in the high-resolution ω/2θ scan [the inset in Fig. 2 (a)]. The film thickness calculated according to the peak position and the fringes is 8.31 nm (*i.e.*,38 u.c.), which is consistent with the calibrated growth rate.

To obtain further structural information about the LCO/STO(111) sample, we performed XRR measurements by using a grazing incidence X-ray reflectometer in the θ-2θ mode. The experimental curve, with seven well-defined narrow Bragg peaks observable up to 4° [black line in Fig. 2 (b)], indicates the high-quality epitaxial growth with the smooth surface, well-defined LCO crystallinity, as well as the sharp LCO/STO heterointerface. The XRR curve was used to determine the film thickness and interfacial roughness. The position of the Bragg peaks corresponds to a film thickness of 8.25 nm, which is in excellent agreement with the results calculated from XRD measurements and the calibrated growth rate. The deduced interfacial roughness, including the contributions from both geometrical roughness and interdiffusion, is about 0.37 nm.

The chemical states of the sample surface were investigated by *ex-situ* XPS measurements. As seen in Fig. 2 (c), the Co 2p core level shows the multiplet splitting of 15.2 eV between the Co $2p_{3/2}$ and Co $2p_{1/2}$ peaks (binding energies of 779.9 eV and 795.1 eV, respectively), which is a typical characteristic of the $Co^{3+}$ ions [31]. Note, in the case of $Co^{2+}$ containing compounds the binding energy separation is usually around 15.8 eV. We also recorded the core levels of La and O, which show a slight variation in the O relative surface concentrations due to the variations in the surface contamination level. However, the La/Co/O chemical content is nearly 1:1:3, as expected for $LaCoO_3$. Since XPS only probes the near surface layers of the sample, synchrotron-based resonant XAS (TEY mode with penetration depth of 2-10 nm) at Co $L_{2,3}$-edge was performed to further confirm the right valence states and local environments of $Co^{3+}$ cations. As shown in Fig. 2 (d), the XAS spectra at Co $L_{2,3}$-edge with two dominative features near 780 eV ($L_3$) and 796 eV ($L_2$) is a conventional spectra of LCO agreeing very well with reported $Co^{3+}$ based compounds [32].

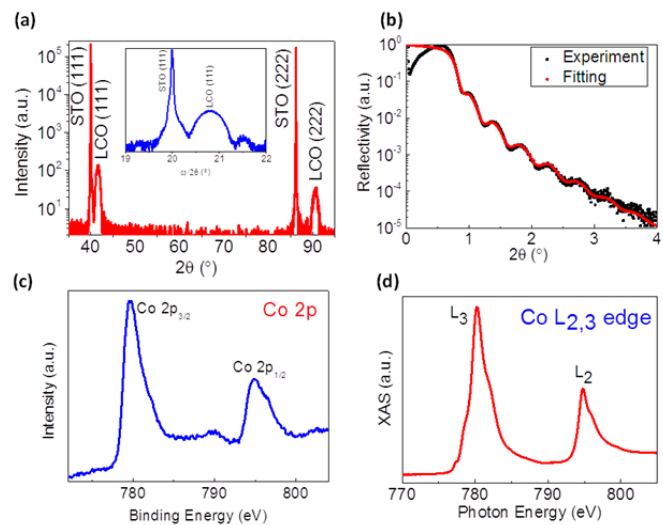

Figure 2 (a) XRD of LCO film on STO(111). (b) XRR and the fittings. (c) XPS for Co 2p core level. (d) XAS spectra at Co $L_{2,3}$-edge fitted with TEY model at room temperature.

Next we discuss the superlattices of LCO/STO(111). We employed the same interrupted deposition method to grow the LCO/STO superlattices on STO(111) substrates. When STO is deposited on LCO, the typical half-orders streaks ("2×" reconstructions) became weaker and vanished as the layer coverage reach 1 u.c.. Moreover, the RHEED patterns revealed the $SrO_2$-terminated surface with (4×4) reconstruction after

STO layers growth [the inset in Fig. 3 (a)]. Further deposition of LCO on STO precise recovers the "2x" reconstruction on the surface. This growth sequence was repeated with the respective thickness control so that the high quality LCO/STO superlattices were obtained. In the following we discuss the $[(LCO)_{10}/(STO)_{20}]_3$ superlattice as a representative example. The cross-sectional STEM imaging in the high-angle annular dark-field (HAADF) mode along $[1\bar{1}0]$ direction clearly shows the excellent crystallinity of the LCO and STO layers, their sharp interfaces and the well-ordered superlattice arrangement of layers, as displayed in Figs. 3 (a) and (b). Since the image contrast of the HAADF-STEM micrograph is related to the elemental concentration, a line profile is drawn along the A-site cations (La/Sr) across the LCO/STO interface [Fig. 3 (c)]. As seen, La and Sr shows a slight intermixing with the length scale of less than 2 u.c. (~ 0.38 nm), which is in excellent agreement with the result calculated from the XRR measurement. The STEM imaging confirmed that the LCO/STO superlattice has sharp interfaces in the [111]-oriented oxide heterostructures.

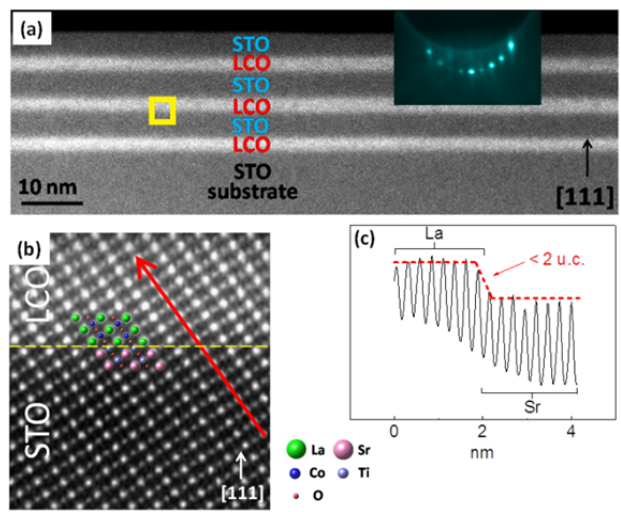

Figure 3(a) HAADF-STEM image of $[(LCO)_{10}/(STO)_{20}]_3$ superlattice grown on STO(111) substrate. The inset shows the RHEED pattern of the grown surface with (4×4) reconstruction. (b) HADDF-STEM image of the yellow rectangular area in (a). (c) Line profile of the image brightness at La/Sr sites along the red arrow in (b).

In summary, we synthesized and characterized the LCO(111) thin films and [111]-oriented

LCO/STO superlattices on STO substrates in 2D mode by PLD. The STO substrate was treated to form a reconstruction for the atomically flat surface which was preserved during the growth. Interrupted deposition at a lowered growth temperature was used to suppress the chemical diffusion across the heterointerfaces. Comprehensive characterizations confirmed the high crystallinity, proper chemical valence, and sample uniformity of the LCO layers, as well as the interface sharpness of the superlattices. The current work demonstrates a path to grow designed digital superlattices even in the presence of strong surface polarity and interface polar discontinuity, thus paving a way to explore the rich quantum phenomena in oxide heterostructures.

We gratefully acknowledge Alpha N'Diaye for the support and assistance of XAS measurement. This work was supported by the National Natural Science Foundation of China (No. 11474334 & 11634016) and the National Key Research & Development Program of China (No. 2017YFA0303600). M.-H. Hu acknowledges the China Postdoctoral Science Foundation (No. 2015M571147) and the Scientific Research Foundation for the Returned Overseas Chinese Scholars. Y.C. and J.C. were supported by the Gordon and Betty Moore Foundation's EPiQS Initiative through Grant GBMF4534. MK was supported by the Department of Energy under grant DE-SC0012375.

Zhang, Science 318, 766 (2007).

[6] Y. Xia, D. Qian, D. Hsieh, L. Wray, A. Pal, H. Lin, A. Bansil, D. Grauer, Y. S. Hor, R. J. Cava, and M. Z. Hasan, Nat. Phys. 5, 398 (2009).

[7] Y. L. Chen, J. G. Analytis, J.-H. Chu, Z. K. Liu, S.-K. Mo, X. L. Qi, H. J. Zhang, D. H. Lu, X. Dai, Z. Fang, S. C. Zhang, I. R. Fisher, Z. Hussain, and Z.-X. Shen, Science 325, 178 (2009).

[8] N. Nakagawa, H. Y. Hwang, and D. A. Muller, Nature Mater. 5, 204 (2006).

[9] A. Savoia, D. Paparo, P. Perna, Z. Ristic, M. Salluzzo, F. MilettoGranozio, U. Scotti di Uccio, C. Richter, S. Thiel, J. Mannhart, and L. Marrucci, Phys. Rev. B 80, 075110 (2009).

[10] Y. Hotta, T. Susaki, and H. Y. Hwang, Phys. Rev. Lett. 99, 236805 (2007).

[11] M. Takizawa, Y. Hotta, T. Susaki, Y. Ishida, H. Wadati, Y. Takata, K. Horiba, M. Matsunami, S. Shin, M. Yabashi, K. Tamasaku, Y. Nishino, T. Ishikawa, A. Fujimori, and H. Y. Hwang, Phys. Rev. Lett. 102, 236401 (2009).

[12] J. Blok, X. Wan, G. Koster, D. Blank, and G. Rijnders, Appl. Phys. Lett. 99, 151917 (2011).

[13] S. Middey, D. Meyers, D. Doennig, M. Kareev, X. Liu, Y. Cao, P. J. Ryan, R. Pentcheva, J. W. Freeland, and J. Chakhalian, Phys. Rev. Lett. 116, 056801 (2016).

[14] M. Saghayezhian, Zhen Wang, Hangwen Guo, Yimei Zhu, E. W. Plummer, and Jiandi Zhang, PHYSICAL REVIEW B 95, 165434 (2017).

[15] P. M. Raccah and J. B. Goodenough, Phys. Rev. 155, 932 (1967).

[16] M. A. Señarís-Rodríguez, and J. B. Goodenough, J. Solid State Chem. 116, 224 (1995).

[17] T. Saitoh, T. Mizokawa, A. Fujimori, M. Abbate, Y. Takeda, and M. Takano, Phys. Rev. B 55, 4257 (1997).

[18] K. Asai, A. Yoneda, O. Yokokura, J. M. Tranquada, G. Shirane, and K. Kohn, J. Phys. Soc. Japan 67, 290 (1998).11